\documentclass[12pt]{article}

\usepackage{mathrsfs,amsbsy,amssymb,latexsym,amsfonts,amsmath,cite}

\global\arraycolsep=3pt
\newcommand\CL{\mathcal{L}}

\newcommand\DS{\mathscr{D}}
\newcommand\CO{\mathcal{O}}
\newcommand\CS{\mathscr{C}}

\newcommand\e{\mathrm{e}}

\newcommand\Ds{D\!\!\!/} 
\newcommand\lhc{LHC }

\newcommand\path{\mathrm{path}}

\renewcommand\chi{r}
\renewcommand\xi{a}
\renewcommand\kappa{p}

\begin{document}

\begin{flushright}
	YITP-07-45\\
	OIQP-07-08 
\end{flushright}

~
\vspace{20mm}

\begin{center}
{\Large 
Search for Effect of Influence from Future \\in Large Hadron Collider 
}

\vspace{20mm}

{\large
Holger B. Nielsen}
{\it
\footnote{
On leave of absence to CERN, Geneva from 1 Aug. 2007 to 31 May 2008.}\\
The Niels Bohr Institute,
University of Copenhagen,
\\
Copenhagen $\phi$, DK2100, Denmark
}
\\
and
\\
{\large
Masao Ninomiya}
{\it
\footnote{
Also working at Okayama Institute for Quantum
Physics, Kyoyama 1, Okayama 700-0015, Japan.}\\ 
Yukawa Institute for Theoretical Physics,\\
Kyoto University, Kyoto 606-8502, Japan}\\

\end{center}

\vfill
\begin{abstract}

We propose an experiment which consists of drawing a card and using
it to decide restrictions on the running of Large Hadron Collider (LHC for short) at CERN, 
such as luminosity, and beam energy.
There may potentially occur total shut down.
The purpose of such an experiment is to search for influence from
the future, that is, backward causation.
Since LHC will produce particles of a mathematically new type of
fundamental scalars, i.e., the Higgs particles, there is potentially
a chance to find 
unseen effects, such as on 
influence going from future to past,
which we suggest in the present paper.

\end{abstract}
\newpage

\section{Introduction}

In general, it is believed, because of causality, that backward causation \cite{1}, 
in the sense that what happens at a later time influences 
what happens earlier, does not occur.

However each time we surpass a new energy scale so as to produce, for 
example, a type of particle with new mathematical properties,
we should retest our well-working principles of earlier experiments.

This model of ours is a model of the initial conditions
of the Universe.
It may be viewed as having a similar condition
to the ``no-boundary"
initial condition postulated by Hartle and Hawking \cite{2}
at the moment of the birth of the Universe.

Our theoretical model building \cite{3,4,5}, 
in particular, calls for a retest.
When the Higgs particles are to be produced,
we must carry out a retest to elucidate whether 
there could be an influence from the future.
For instance, the potential production of a large number of 
Higgs particles at a certain future time would cause a
pre-arrangement such that Higgs particle production can be avoided.
Such pre-arrangements may be considered an influence from the future.
One of us (H.B.N.) has contemplated, through the past several years, 
the idea of an influence from the future on the other settings
\cite{6,7}.
One also finds such future influences on effective coupling constants
in ``Baby Universe Theories"
\cite{8,9,10,11,12,14},
and in some models of the ``multiple point principle"
\cite{7,13,14}.

%%%%%%%%%%%%%%%<•ÊŽ†ƒy[ƒW'Q'ð'}"ü'·'é>%%%%%%
In section 2 we review our model used in the present
article of which action consists of real and imaginary parts.
In section 3 we propose the experiment at \lhc to verify some effects
advocated by us.
Section 4 is devoted to estimating the probabilities of various
cases predicted by card game experiment.
In section 5, we check our model numerically.
In section 6 we investigate Higgs particles in terms of 
particle and field actions.
In section 7 we explain why we consider the Higgs particle to be so special.
In section 8, we estimate how the width of the Higgs particle
is expected to be broadening.
The last section 9 is devoted to the conclusions and outlook.

\section{Our model with imaginary part of action}
In our previous publications \cite{3,4,5}
we described our model by simply introducing 
a functional called $P[\path]$ depending on the path, 
which could be most easily thought of as a classical path of all the 
fields in the universe, 
and $P[\path]$ denotes the probability that this path is realized. 
The idea behind $P$ should be that it is calculable from some 
physically reasonable formula involving the path, 
since we would like to let $P$ depend on the path in such a way that it obeys
the usual physical symmetries and principle of locality in space time.
Thus it is expected to be such a form as 
\begin{eqnarray}
P \simeq \e^{-2S_I[\path]},
\end{eqnarray}
where $S_I[\path]$ is the action
given by the imaginary part of the Lagrangian $\CL_I$,
\begin{eqnarray}\label{1.2}
S_I[\path]=\int \CL_I(x)\sqrt{g} d^4x
.\end{eqnarray}

The formulation of our model is such as to simply allow the
action $S[\path]$ in the Feynman path way integral
$\int \e^{i S[\path]}\DS[\path]$ to be complex, 
\begin{eqnarray}\label{1.3}
S[\path]= S_R[\path]+iS_I[\path],
\end{eqnarray}
and then assume that the imaginary part of the Lagrangian density
$\CL_I(x)$ in (\ref{1.2}) 
has much the same form as the real part $\CL_R(x)$ 
in, for example, the Standard Model 
or some extension thereof.
The only difference between $\CL_R$ and $\CL_I$ is
the coefficients, such as $\frac{1}{g_N^2}, Z$ and $m^2$ 
of the various terms in (\ref{1.3}): 
\begin{eqnarray}
-\frac{1}{g_N^2}F_{\mu\nu}^a(x)F^{a\mu\nu}(x)~,~~~
Z\bar\psi\Ds\psi, \cdots ,
,\end{eqnarray}
However 
the forms of the field dependences are the same,
since the renormalization factors Z and other coupling constants, 
are different in $\CL_R$ and $\CL_I$.
At first sight a model of this type seems to be obviously false,
since $S_I[\path]$, which gives the probability of the development 
of a path
of the universe, would a priori depend strongly
on what goes on today or at a later time.
Such effects would appear as if 
that the universe were prearranged to achieve various goals 
that would be obtained by the largest possible negative contributions to $S_I$.
However, we believe we have found some arguments that the 
importance of the inflation era should be much more than the 
present era in selecting the path to be realized, and henceforth, 
the dependence on what goes on today is strongly suppressed.
In this way, we claim to be able to obtain the second law of 
thermodynamics out from our model.

The mechanism of governing the development of the universe so as to avoid
the production of Higgs particles was suggested in our previous works.

We have already proposed a model for unifying equations of motion 
and the choice of the initial conditions, or better, the selection
of the solutions of the equation of motion to be realized.
It is, at least, some unification to obtain the selection of the solution
to be realized by some law.
The very unusual feature of this type of model is that such an
imaginary part of the action
\begin{eqnarray}
S_I=\int \CL_I dt,
\end{eqnarray}
which leads to the probability weighting
\begin{eqnarray}
\e^{-2 S_I},
\end{eqnarray}
depends not only on the happenings at the very first moment of the 
birth of the Universe, but also on what happens
at all times.
If we did not provide detailed speculations that the main effect
on selecting the solution to be realized is from the big bang era,
our model would be falsified by the upper bound on the occurrence of 
prearranged events, or by the second law of thermodynamics.

We have, however, some rather naturally working mechanisms
\cite{3,4}
that can make the effects of the imaginary part of the action 
negligible under some conditions that likely prevail until LHC
starts colliding beams.
In fact, we have, in earlier articles 
\cite{3,4},
argued that the imaginary part $\CL_I$ of the Lagrangian would be
constant - and thus unimportant - in one of the following three cases \cite{3}.
\begin{enumerate}
  \item[1)]
 If particles are either
\begin{itemize}
  \item[(a)] nonrelativistic and conserved or 
  \item[(b)] massless (photons), 
\end{itemize}
then the $S_I$-effect will be negligible. 
  \item[2)] Even with relativistic particles, the effect of $S_I$
vanishes provided the Lagrangian 
$L_R + iL_I$
is homogeneous in a field type and has only one 
independent coefficient by symmetry restrictions; 
for example, the Standard Model
Lagrangian with $SU(3)\times SU(2)\times U(1)$
is homogeneous in the second order in the quark and lepton fields, 
and we only use one term unless Yukawa couplings are added.
However, the latter can be chirally transformed to have no independent
phase relative to one of the kinetic terms, e.g., the right-handed one.
  \item[3)] In addition, the imaginary part of the Lagrangian $L_I$
for a Yang Mills theory is forbidden provided there exist monopoles
\cite{3,15}.
\end{enumerate}
In daily life, point 1) is sufficient to suppress the effects of 
influences from the future via $L_I$, so that no prearrangements would
occur strongly there.
However, high-energy physics machines dealing with their relativistic
particles would, if it were only for 1), influence their past. 
For instance, such an influence could have meant that these machines
would have met with bad luck by prearrangement, whereby their
funds may have been cut so that they would not be in operation.
Seemingly there were no such effects of bad luck for relativistic
accelerators such as ISR, wherein the particles are even stored for
long times.
To rescue our model, which is already falsified by ISR,
we could, however, make the very mild speculation
that there exists fundamentally magnetic monopoles 
\cite{15}, 
which is allowed for the Yang Mills fields in the Standard Model.
Such an existence of monopoles, together with the remark that the Lagrangian of the
fermions - quarks and leptones - is homogeneous in the fermion
fields, could provide, by means of 2) and 3), the argument for the fact that, 
even for the high energy experiments performed so far, no effects of 
bad or good luck have been observed.

However, the Higgs particles are the first fundamental scalar to be
investigated and arguments 1), 2), and 3) above may very likely 
be insufficient for eliminating 
the effect of influences from the future related to Higgs particles.

Thus it is likely that our expectation of our model in the paper 
``Influence from future..."\cite{3} might show up in the same
experiment that first produce big amounts of Higgs particles.

Very interestingly, in this connection, the SSC in Texas
\cite{16}
would have been the first machine to produce Higgs particles 
on a large scale.
However, it's construction was actually stopped after one-quarter of
the tunnel was built, which is almost a \underline{remarkable} piece of
bad luck.

\section{Proposal for the experiment}

If we just, in a very general way, consider a model in which the probability
$P(sol)= \e^{-2S_I(sol)}$
for a solution, denoted by $sol$, of the equations of motion
to be realized is a function of what
happens to this solution $sol$ at all different times $t$, we should
be able to see influences from the future.
If, as is suggested above, $P(sol)$ depends on whether or not Higgs
particles are
produced in large amounts during the development of the world $sol$,
then the actually realized development would either seek
or avoid Higgs particle production.
It seems most likely that the production of Higgs particles leads 
to smaller $P(sol)$ than that for no Higgs particle production,
 since otherwise, there
would already have been many Higgs particles produced in nature.

With this model, we expect that a 
machine for producing Higgs particles will be stopped by some accident
or another
if the effect is sufficiently large with 
having in mind that the probability should exponentially decrease with the 
number of Higgs particles produced.
The ratio of the two probabilities, 
\begin{eqnarray}
\frac{P(sol_\mathrm{with~machine})}{P(sol_\mathrm{without})}
\sim
C^{\sharp\mathrm{Higgses}},
\end{eqnarray}
may be obtained.
Here, $sol_\mathrm{with~machine}$ and $sol_\mathrm{without}$ indicate the
solution with and without the machine, respectively.

The experiment proposed in the present article is to give 
``foresight", a chance of avoiding forced closure of 
LHC due to lack of funding or other form of bad luck, as happened
to SSC.

We imagine a big stack of cards on which 
are written various restrictions concerning the operation of LHC, 
for example ``allow the production of only 10 Higgs particles".
On most of the cards there should just be written ``use LHC freely"
so that they cause no restrictions.
However, on a very small fraction of cards, there should be restrictions
on luminosity or beam energy or some combination of them.
One card may even have ``close (shut down) LHC".

The crucial idea of this proposal is that if our model were true, 
then the most likely development $sol$ with the 
$P(sol)\simeq  \e^{-2S_I(sol)}$ factor included would be 
a development involving one of
the cards which strongly restricts on the Higgs particle production at LHC.

\section{Estimation of probabilities of each case:
Probability of closing LHC}

Before setting down the rules of the card game, 
one should carefully discuss what is the most economical and
optimal probability value to choose for, for instance, the ``close LHC card".

In order to give an idea about what probability $\kappa$ to choose for 
closing LHC,
while postponing partial closings or milder restrictions until the next
section,
we shall introduce the following symbols for the relevant probabilities.

$\chi$: the probability that our model is correct so that there is a
prearrangement mechanism ensuring that LHC will not come into operation.

$\xi$: the probability that without any such mysterious interference, 
the LHC will accidentally fail and thus not start.

$d$: the average excess damage that occurs under an accidental bad-luck
event preventing LHC from working; it may be a larger value than that of LHC itself.

$\kappa$: the probability value for the ``close LHC" card 

The numbers $\chi, \xi$ and $\kappa$ should be very small,
whereas the excess average damage, is presumably of order unity.
One could, however, estimate that this extra damage involves
even human lives.
Thus several people may be killed during some explosion.
In such a case the damage could turn out to be more severe than
the pure loss of LHC itself.
Hence we might take the probability $d$ to be one order of
magnitude larger than the value for LHC.
A reasonable value may be 
$d\approx 10$.

In the case that probabilities $r, a$, and $p$ are all small LHC will
most likely come to work as expected without any problem. There will
only be the small probability a that it has a normal accident and
the small probability $p$ that it gets closed due to the card game proposed.
In the case of probability $\chi$, \lhc
cannot be allowed to start up.
It can fail in two ways: 
with probability
$\frac{\xi}{\xi+\kappa}\cdot\chi$, 
there will be a normal accident, and extra damage may be 
given by the factor $d$; 
with probability 
$\frac{\kappa}{\xi+\kappa}\cdot\chi$, 
\lhc will be stopped by the card ``close LHC".
To estimate these probabilities, we considered 
that the two types of stoppage should occur with a relative probability
$\xi:\kappa$, 
as also if our theory were wrong.

We can now estimate the average cost due to the various failures
in the natural units of the value of LHC.

Let us denote by $\CS$ the average loss due to severe failure
in units of the price, for example, $3.2 \times 10^9$ CHF (Swiss Francs), 
of \lhc itself.
\begin{eqnarray}\label{8}
\CS=
\kappa+\xi\cdot(d+1)
+\chi\left(
\frac{\kappa}{\xi+\kappa}+\frac{\xi}{\xi+\kappa}\cdot(d+1)
\right)
\end{eqnarray}
Here, we took the loss, $(d+1)$, by natural failure 
as the sum of excess loss $d$ and loss, 1, of the machine itself.
Simplifying (\ref{8}), we get
\begin{eqnarray}
\CS=
\left(\kappa+\xi\cdot(d+1)\right)
\left(
1+\frac{\chi}{\xi+\kappa}
\right)
~.\end{eqnarray}
Since $\kappa$ is at our disposal, one would say that we should
choose it on the basis of ethical and economical reasons so as to minimize
the loss in \lhc price units.
This minimization occurs for the case, 
\begin{eqnarray}
\frac{\partial\CS}{\partial\kappa}&=&
(1+\frac{\chi}{\xi+\kappa})
+(\kappa+\xi(d+1))
(\frac{-\chi}{(\xi+\kappa)^2})\nonumber\\&=&
1-\frac{\xi\chi d}{(\xi+\kappa)^2}
\nonumber\\&=&0,
\end{eqnarray}
which leads to
\begin{eqnarray}
\kappa^2+2\kappa\xi+\xi^2-\xi\chi d=0.
\end{eqnarray}
The solutions are given by
\begin{eqnarray}
\kappa=-\xi\pm\sqrt{\xi\chi d}~.
\end{eqnarray}
Of course, we must account for the chance that the closing card $\kappa$
is non-negative, $\kappa\ge 0$.

If $\sqrt{\xi\chi d}<\xi$, 
\begin{eqnarray}
\chi < \frac{\xi}{d},
\end{eqnarray}
i.e., if the chance of our theory being right, $\chi$, is less than
the chance of a natural failure of \lhc divided by the excess damage
factor $d$, then it would not be optimal to play our card game for any
possible value.
It would cause damage to perform our experiment
and one should only do it in order to confirm
(or invalidate) our theory.

If, however, one judges that the chance of our model being corrected 
is so large that
\begin{eqnarray}\label{relevant}
\chi > \frac{\xi}{d},
\end{eqnarray}
then, it would be uneconomic and unethical not to perform our card
game.

In this case, supposing that we only compute the
optimal value of $\kappa$ as orders of magnitude 
in order to avoid damage,, the value should be 
\begin{eqnarray}
\kappa=-\xi+\sqrt{\xi\chi d}
\approx \sqrt{\xi\chi d}~.
\end{eqnarray}

This obviously means the following. 
Unless the chance of drawing the card ``close LHC", $\kappa$, is at least
as big order of magnitude as the chance of a normal failure
of LHC, 
it is ineffective, for preventing damage, to play our
proposed card game.
This means that, if we choose $\kappa \ll \xi$,
even if our theory were right, \lhc
would be stopped by a normal failure rather than by our card game.

\section{Consideration on checking our model}

The purpose of playing the proposed card game is 
to carry out a very clear test of our model
in addition to an economical or
ethical attempt to rescue \lhc from even worse fates.
Crudely speaking, a superficially ``normal" accident
would already be strong support for our model.
However, it would be even more valid numerically if LHC were stopped
by a card play.
Then one would have a very clear knowledge of the statistical
accuracy with which our prearrangement effect had worked and had
been tested.
To know in advance a good estimate for probability $\xi$ 
is not so easy.
Therefore one could reason away such a natural failure and
say that, in spite of it, one should not trust our theory.
One could say
``oh, it is an accident of bad diplomacy".
Drawing a single specific card from among 2 million cards
could only be achieved either by a card magician or by a model like
ours.
In principle, such an unlikely occurrence would be possible 
but not in practice!

In order for our model to be safely confirmed, we must choose
$\kappa$ to be so small that drawing the card ``close LHC" would indeed be
convincing.
To suggest a value for $\kappa$, we recall that the discovery
of the Higgs particle is suggested to be performed with a 5-standard-deviation 
peak.
A 5-standard-deviation peak occurs by accident in a band only with the probability
$5 \times 10^{-7}$, i.e., one in 2 million.
If one were to trust these 5-standard-deviation discoveries even 
if one had, for example, 
10 mass bands in which to look for the Higgs peak, it would mean
that one would accept a discovery even if an accidental reproduction
of data were to occur with the probability 
$10 \times 5 \times 10^{-7}= 5 \times 10^{-6}$
or one in 200,000.

An experiment of our proposed concept with a probability for stopping
LHC of $p \approx 5\times 10^{-6}$ would mean on average an expense
equal to $5\times 10^{-6} \times $ ``cost of LHC"$=5 \times 10^{-6}
\times 3.3 \times 10^9 \mbox{ CHF} \approx 1.7 \times 10^4 
\mbox{ CHF}$.
However, this average loss of 17,000 CHF would be compensated by the
danger of a natural stoppage due to explosion or bankruptcy of CERN
or other similar things caused by the effect of $S_I$
in the case that our theory were valid.
Compensation of the average loss would occur if the following is satisfied:
\begin{enumerate}
  \item it is less possible than $\kappa=5 \times 10^{-6}$ that
\lhc would experience a normal failure 

and 
  \item $d$, the ``excess loss", times the chance that our model is 
true, $\chi$, i.e., $d \times \chi$, is larger than $\kappa \sim 5 \times 
10^{-6}$.
\end{enumerate}

If you include the danger that the failure of \lhc
could be due to war between the member states of CERN, the extra
damage $d$ could be very large, but that sounds exaggerated.
Presumably, we should take $d \sim 1$ to 10, for example, 5. Then
\begin{eqnarray}
d \times \chi\sim 5\chi\sim 5\times 10^{-6}=\kappa_\mathrm{suggest}
\end{eqnarray}
for $\chi\sim 10^{-6}$.

In other words, if there were just one chance in a million that our model were
right and if normal failure were extremely seldom, then the
17,000 CHF would already be paid for.

If our model has more than one chance of being right 
in one million, one might rather
begin to worry that taking only $\kappa \sim 5 \times 10^{-6}$
might result in too great a danger. However, if 
this failure by itself $\xi$ is bigger than $5 \times 10^{-6}$, 
our card experiment would fail in the sense that the card drawn 
would not be ``close LHC" even if our model were true.
Nevertheless we might believe our model in that case 
upon witnessing a natural accidental stopping of LHC.
However, that would be less clearly convincing than our card game, and it could
be appreciably more expensive.

We believe that it is essential to perform an honest estimate of the 
reliability of the \lhc construction being completed such that 
the machine is operable, i.e., to estimate probability $\xi$.
Such an estimate of $\xi$ could 
be crucial for the decision as to what value of parameter 
$\kappa$ to choose, i.e., the rules of the card game concerning LHC.

%\section*{Partial closings}
It would presumably pay to make not only the single card 
``close LHC" possibly be drawn, but also to include several cards 
specifying incomplete
closings in the deck.

There could be many variants of the restriction cards, for example, limits on 
beam energy, luminosity and the lifetime of the
machine, and postponement of the start of operation. 
However, all the cards of strong restriction should have only 
low probability $\kappa_0$.
Therefore, just pulling one of them should convincingly confirm the truth
of our model.

\section{Particle action from field action}

In this article, we suggest that the Higgs particle, which
we have not yet studied well, will lead to an influence from 
future effects while 
such an effect is not present for the particles already found:
quarks, leptones, and gauge particles.

In order to explain the peculiarity of the Higgs particle, we shall here 
study the action for a classical particle approximation
to a field theory.
In the usual case of a real action for the field theory, 
one can identify particles as wave packets moving along in the field.
Then, the action one should use is for a particle propagating
in space time,
\begin{eqnarray}
S_\mathrm{R~part}=
2\pi \> \sharp\mathrm{``wave~oscillations"},
\end{eqnarray}
where these wave oscillations are the phase rotations in the wave packet
represented on the field propagating in space time.

It may be a little surprising that the action for the particle 
description $S_\mathrm{R~part}$ is \underline{not}
simply equal to the action contributing to the field theory action $S_R$
from the wave packet.

Indeed it is easy to see that if the Lagrangian 
with respect to a certain
type of field is homogeneous, for instance the $\psi$-involving
part of the Lagrangian
\begin{eqnarray}
\bar\psi(\Ds-m+g_y\varphi)\psi ,
\end{eqnarray}
then the Lagrangian can be constructed from the equation of motion
\begin{eqnarray}
(\Ds-m+g_y\varphi)\psi=0.
\end{eqnarray}
It follows that in the classical field approximation, the action for 
$\psi$ is zero (on the shell).
Hence it is necessary that the effective action for the particle
description is \underline{not} simply the contribution to the 
field theory action $S_R$, because then, we would have only zero
contribution from all the free particles
(in between interaction points).

We have already seen that the main physical significance of the 
imaginary part $S_I$ of the action is that a path under development
is assigned the probability 
$P \sim e^{-2S_I}$ 
so that $S_I$ has the meaning minus half the logarithm of 
the probability weight.
In shifting from the field description to the particle description with
particles, the ``wave weight" $P$ and thus
$S_I=-\frac{1}{2}\log P$
should have the same meaning if we describe the same development 
in the two different languages.
Thus, contrary to what we just claimed for the real part $S_R$, that
\begin{eqnarray}
S_\mathrm{R\, part}\neq S_R,
\end{eqnarray}
we need, for the imaginary part 
- due to its physical significance -
to have the correspondence
\begin{eqnarray}
S_\mathrm{I\, part}= S_I
.\end{eqnarray}
However, it is easily seen that the argument for vanishing action 
in the homogeneous case works to make both 
real $S_R$ and imaginary $S_I$ and 
parts of the actions zero.
Because phenomenologically we do not see any prearrangement
effects involving quarks and leptones,
we must take this to mean that in the particle description,
$S_\mathrm{I\, part}=0$
for particles described by the homogeneous action.

\section{What is so special about the Higgs particles?}

The special property of the Higgs particle that makes it such a favourite
candidate for showing the effects due to our imaginary part of
action $S_I$ is that 
1) it is not a gauge particle and so 
the argument of the nonexistence of the monopole
can be used to exclude imaginary coefficients, and 
2) in the free part of the Higgs Lagrangian, there are two terms,
the kinetic term $|D_\mu\varphi_H|^2$ and the mass term $m^2|\varphi_H|^2$,
of which coefficients have been unrestricted by symmetries, so that these
independent coefficients could have different phases.

Also, for quarks and leptones, one has, at first, independent coefficients
on the kinetic and mass terms, but 
for them, one can perform the chiral transformation
\begin{eqnarray}
\psi_L &\to & \psi_L, \\ \nonumber
\psi_R &\to & e^{-i\delta} \psi_R, 
\end{eqnarray}
which can be adjusted such that the mass and the kinetic coefficients 
have the same phase.
Thereby, the imaginary part of Lagrangian 
$\CL_{I~ \mathrm{quarks ~\&~ leptons}}$ is forced to be proportional to the
real part $\CL_{R~ \mathrm{quarks ~\&~ leptons}}$.
Since the Lagrangians are homogeneous of the second order, one gets
$\CL_{R~ \mathrm{quarks ~\&~ leptons}}=0$ and 
$\CL_{I~ \mathrm{quarks ~\&~ leptons}}=0$ using equations of motion.
However, the Higgs Lagrangian, 
\begin{eqnarray}
\CL_\mathrm{Higgs}(x)=
Z|\partial _\mu\varphi_H|^2
-m^2|\varphi_H|^2
-\frac{\lambda}{4}|\varphi_H|^4,
\end{eqnarray}
is not homogeneous because of the $\frac{\lambda}{4}|\varphi_H|^4$ term,
which is of the fourth order, contrary to the rest. This
could be a further reason for the lack of an argument for
$\CL_{I~ \mathrm{Higgs}}$ to vanish.
The equations of motion,
\begin{eqnarray}
Z\partial _\mu \partial ^\mu \varphi_H
-m^2\varphi_H
-\frac{2\lambda}{4}|\varphi_H|^2\varphi_H=0
\end{eqnarray}
and 
\begin{eqnarray}
Z\partial _\mu \partial ^\mu \varphi_H^\dag
-m^2\varphi_H^\dag
-\frac{2\lambda}{4}|\varphi_H|^2\varphi_H^\dag=0,
\end{eqnarray}
obtained by multiplication with fields 
$\varphi_H^\dag$ and $\varphi_H$, respectively, and adding and subtracting,
does not lead to 
both real and imaginary parts of the field theory action being zero. 
Rather the Lagrangian on shell values are given by
\begin{eqnarray}
\CL_R&=&-\frac{\lambda_R}{4}|\varphi_H|^4~,\nonumber\\
\CL_I&=&-\frac{\lambda_I}{4}|\varphi_H|^4
~.
\end{eqnarray}
Here, we have defined the self-coupling of $\varphi_H$, $\lambda$,
as the sum of real and imaginary parts:
\begin{eqnarray}
\lambda=\lambda_R+i\lambda_I.
\end{eqnarray}
For the Higgs field, one should keep in mind that there is a big
background or vacuum expectation value $\langle\varphi_H\rangle$.
It is, in fact, only the \textit{extra} 
contribution coming from a true particle that is propagating through this
vacuum and described by a wave packet in $\varphi_H$.

We may consider a single Higgs particle described by a wave packet
in the Higgs field $\varphi_{Hwp}$.
Then we obtain, in the well-known background field case with
$\langle\varphi_H\rangle=\varphi_{H bg}=$ 246 GeV$/\sqrt{2}$, 
\begin{eqnarray}
|\varphi_H|^2\approx |\varphi_{Hbg}|^4+|\varphi_{Hbg}|^2
\cdot4|\varphi_{Hbg}|^2+\cdots~.
\end{eqnarray}
This means that we
get a contribution to $S_I$ which again is identified with the
particle $S_\mathrm{I\,part}$ given as 
\begin{eqnarray}
S_\mathrm{I\, part}=
-\frac{\lambda_I}{4}|\varphi_{Hbg}|^2\int
|\varphi_{Hwp}|^2d^4x
~.\end{eqnarray}
The density in 3-space of genuine Higgs particles with energy $E_H$ is
\begin{eqnarray}
\rho=\varphi^*_{Hwp}\overleftarrow{\partial}\varphi_{Hwp}
\sim|\varphi_{Hwp}|^2\cdot E_H,
\end{eqnarray}
so that 
\begin{eqnarray}
S_\mathrm{I\,part}=
-\frac{\lambda_I}{4}|\varphi_{Hbg}|^2
\iint\frac{1}{E_H}\rho d^3\vec{x} dt
~.\end{eqnarray}
For one particle, we have the normalization
\begin{eqnarray}
\int \rho d^3\vec{x} =1~.
\end{eqnarray}
For Higgs particles with reasonably well-defined energy 
$E_H$, the eigentime $\tau$ differential
\begin{eqnarray}
d\tau=
\frac{m_H}{E_H} dt,
\end{eqnarray}
and thus we simply get
\begin{eqnarray}
S_\mathrm{I\,part}=
-\frac{\lambda_I}{4}|\varphi_{Hbg}|^2
\frac{1}{m_H}\int d\tau
~.\end{eqnarray}
Therefore, the imaginary action in terms of Higgs particles is, as expected, 
the eigentime integral $\int d \tau$ multiplied by the 
constant $-\lambda_I|\varphi_{Hbg}|^2\frac{1}{m_H}$.
We do not truly know the imaginary part $\lambda_I$ of the
self-coupling of Higgs particles, 
but a priori, the guess for it would be the dimensionless
of order unity, or rather, the same order as the real part $\lambda_R$
which is of the order $\frac{1}{3}$, for example.

\section{Estimation of effect of Higgs particle}

We see that the contribution to $S_I$ from a Higgs particle seen
from its rest frame with the lifetime $\tau_\ell$, i.e., with
\begin{eqnarray}
\int^\mathrm{decay}_\mathrm{production}d\tau 
=\tau_\ell ,
\end{eqnarray}
is
\begin{eqnarray}
S_\mathrm{I\,part}=-\frac{\lambda_I}{4}|\varphi_{Hbg}|^2
\cdot
\frac{\tau_\ell}{m_H}~.
\end{eqnarray}

Even if we set the Higgs width \cite{17} as large as 1 GeV, for example, 
the order
of magnitude of the exponent in
the decreasing factor of the probability becomes of the order of 100.
The exponentiated value of this becomes so large that no
Higgs particles would be allowed to achieve  so long a lifetime.
Rather, we should expect the Higgs particles to be
brought to decay much faster ``by prearranged accident".
We expect an effective allowed width to be of the order
\begin{eqnarray}
\frac{1}{\tau_\ell}\approx 
\frac{2\lambda_I|\varphi_{Hbg}|^2}{m_H}
~.\end{eqnarray}

Looking for this broadening of the Higgs width according to the
effect of our model might be in itself a very interesting 
prediction
\cite{3}.
However, once the broadening takes place, the effective 
``decreasing factor of the probability" 
will only be of the order of unity or at least no smaller
than of the order
$\frac{\Gamma_\mathrm{apriori}}{2\lambda_I|\varphi_{Hbg}|^2/m_H}$.
This would mean of the order of a factor of 
$\frac{1}{\CO(100)}$
rather than
$\e^{-\CO(100)}$.
Because of such a mechanism of making Higgs particle decay ``miraculously"
fast, the suppression by each Higgs particle by more than the order of 
unity may be avoided.
Thus a few Higgs particles might be allowed, as may 
already have been seen at LEP, but huge
amounts of Higgs particles should be completely avoided.
Machines such as LHC, which makes many Higgs particles should be
stopped quickly, before having made more than a few Higgs particles!

Particles other than the Higgs particles lack the self-interaction 
term of the fourth order, except in the 
Yang-Mills field.
Indeed because of renormalizability requirements, the Fermion fields
$\psi$ and $\bar{\psi}$ for quarks and leptones, respectively, cannot be allowed to be
more than second order.
For them, therefore, $S_\mathrm{I\, part}=0$, while we seek to suppress 
the Yang-Mills contribution to imaginary $\CL_I$ on the basis of 
our argument about assuming monopoles
\cite{3,15}
\footnote{If we assume that our model implies MPP 
\cite{7, 12, 13}
we should get the lowest allowed standard model Higgs mass \cite{18}, 
which allows only b$\bar{\mathrm b}$-decay and a much smaller width by
a factor of 500.}

\section{Conclusion and Outlook}

In the present article, we have proposed an 
experiment at \lhc for determining the effect of an influence 
from the future as 
proposed in our own model.
The best description may be achieved by introducing 
an imaginary part $S_I$ of the action $S$.
The experiment is very primitive in as far as it consists
simply of a card-drawing game arranged so that some severe
restriction on the running of \lhc - essentially closure - is imposed
with a probability $\kappa$ of the order of $5 \times 10^{-6}$.
If indeed a restriction card 
which has such a low probability as $\kappa \sim 5 \times 10^{-6}$ 
were to be drawn, it would essentially mean that our model must be true!
If, however, just a normal card that gives no restriction is drawn, 
our theory would be falsified 
\textit{unless a seemingly accidental stopping of \lhc occurs}!

It must be warned that if our model were true and no such game
about strongly restricting \lhc were played, or if the probability 
$\kappa$ in the game for restricting were too small, then a 
``normal" (seemingly accidental) closure should occur.
This could be potentially more damaging than just the loss of
\lhc itself.
Therefore not performing
(or not performing with sufficiently big $\kappa$) 
our proposed card game could - if our model were correct -
cause considerable danger.

Of course, a priori
- as just a proposed effect to look for -
the chance $\chi$ that such a model is right is very low.
However, we have already published a few papers \cite{3,4,5}
on this
type of backward causation model, and several predictions seem to
be phenomenologically good:
for instance, we can claim to have speculations that may lead to a 
cosmological constant of the same order as that of matter density
\cite{5}.
That is to say, our model is promising with respect to solving the 
cosmological constant problem and the ``why today" problems.
Also, we claim that it is promising for explaining
why there 
should be a bottom in the Hamiltonian \cite{5}.
A further consequence is the principle of many degenerate vacua
\cite{7,12,13} (MPP = multiple point principle), on which
one of us (H.B.N.) has worked for many years with some success.

Finally, let us mention that we are working on an article which suggests
that our type of model may be able to cope
with the measurement problem in quantum mechanics
\cite{19}.
If one wishes to set the eigenvalue of a
measured quantity 
\textit{before the enhancement of the signal in the measurement 
instrument has occurred}
then some sort of backward causation seems to be called for: 
without the signal enhancement, can one really say if it is a genuine
measurement?

We believe that before performing the proposed experiment, we should
carefully discuss and evaluate the most optimal choice of the
rules of the game.
It is most important to choose $\kappa$ that gives the chance
of closure of LHC in the game.

However, in the case of an essential closure, we might 
obtain interesting information about the details of our 
confirmed model if we include many cards with various partial closings, 
for example, how many Higgs particles can we allow \lhc to produce 
before complete closure?

The allowance of such tiny amounts of Higgs particle production and the 
running of 
\lhc could, if our model were true, provide some information on the 
details of our model.
It is presumably very profitable to organize several
possibilities of partial closings.
Such possibilities
might tell us about the size of $\lambda_I$, for example.

\section*{Acknowledgments}
We acknowledge the Niels Bohr Institute (Copenhagen) and 
Yukawa Institute for Theoretical Physics for their hospitality.
This work is supported by Grant-in-Aids for Scientific Research 
on Priority Areas,  Number of Area 763 ``Dynamics of Strings and Fields", 
from the Ministry of Education of Culture, Sports, Science and Technology, 
Japan. We also acknowledge discussions with colleagues, especially John Renner 
Hansen, on the SSC.


\begin{thebibliography}{99}
\bibitem{1}
 J. Faye,
``The Reality of the Future",
Odense University Press.

\bibitem{2}
 J.~B.~Hartle and S.~Hawking,
 Phys.~Rev.~{\bf D28} 2960-2975 (1983).
 
 
\bibitem{3}
 H.~B.~Nielsen and M.~Ninomiya,
``Future Dependent Initial Conditions from Imaginary Part in Lagrangian",
Proceedings of the 9th Workshop 
``What Comes Beyond the Standard Models",
Bled, 16 - 26 September 2006, 
DMFA Zaloznistvo, Ljubljana,
 hep-ph/0612032.

\bibitem{4} H.~B.~Nielsen and M.~Ninomiya, ``Law Behind Second Law of 
Thermodynamics-Unification with Cosmology'', JHEP ,03,057-072 (2006),
 hep-th/0602020.

\bibitem{5}
 H.~B.~Nielsen and M.~Ninomiya,
 ``Unification of Cosmology and Second Law of Thermodynamics:
 Proposal for Solving Cosmological Constant Problem, and Inflation",
Progress of Theoretical Physics, Vol. 116, No. 5 (2006)
  hep-th/0509205, YITP-05-43, OIQP-05-09.

\bibitem{6}
 H.~B.~Nielsen and S.~E.~Rugh, Niels Bohr Institute Activity Report 1995.


\bibitem{7} H. B. Nielsen and C. Froggatt, 
School and Workshops on Elementary Particle Physics,
Corfu, Greece, September 3-24, 1995.

\bibitem{8}
S. Coleman, Nucl. Phys. {\bf B307} 867 (1988).


\bibitem{9}
S. Coleman, Nucl. Phys. {\bf B310} 643 (1988).

\bibitem{10}
T. Banks, Nucl. Phys. {\bf B309} 493 (1988).

\bibitem{11}
S. W. Hawking, Phys. Lett. {\bf 134B} 403 (1984).

\bibitem{12}
Reviews:
H. B. Nielsen and M. Ninomiya,
Lecture Notes of International Symposium on the Theory of Elementary
Particles,
Ahrenshoop, DDR, October 17-21, 1988.

\bibitem{13}
D. L. Bennet and H. B. Nielsen,
Int. Journ. Mod. Phys A9 (1994) 5155-5200.

\bibitem{14}
H. B. Nielsen and M. Ninomiya,
``Degenerate Vacua from Unification of Second Law of Thermodynamics
with Other Laws", hep-th/0701018.

\bibitem{15}H.-M. Chan and S. T. Tsou,
hep-th/9904102,
April 1999,
Int. J. Mod. Phys A14(1999) 2139-2172.


\bibitem{16}J. Mervis and C. Seife,
10,1126/science, 302.5642.38,
New focus:
10 Years after SSC.
Lots of Reasons, But Few Lessons.
 

\bibitem{17}
L. \v{Z}ivkovi\'{c},
Weizmann Institute,
``Measurements of Standard Model Higgs Parameters of Atlas",
July 20, 2003, 
Conferences/2003/Praha 2003\_Lidiya-Higgs-parameters.ppt.

\bibitem{18}
C. D. Froggatt and H. B. Nielsen,
``Standard Model Criticality Prediction Top Mass 173$\pm$5 GeV
and Higgs Mass 135$\pm$9 GeV",
Phys. Lett. B368 (1996) 96-102.

\bibitem{19}
H. B. Nielsen and M. Ninomiya,
paper in preparation.




 
\end{thebibliography}
\end{document}